\documentclass[10pt]{article}

\usepackage{amsmath}
\usepackage{amsfonts}
\usepackage{amssymb}

\begin{document}

\title{\bf Solar system effects in Schwarzschild--de Sitter spacetime}

\author{{\bf Valeria Kagramanova${}^1$, Jutta Kunz${}^2$, and Claus L\"ammerzahl${}^3$} \\
$~$\\
${}^1$ Institute of Nuclear Physics and Ulugh Beg Astronomical Institute, \\
Astronomicheskaya 33, Tashkent 700052, Uzbekistan \\
kavageo5@rambler.ru \\
${}^2$ Institut f\"ur Physik, Universit\"at Oldenburg, Postfach 2503\\
D-26111 Oldenburg, Germany\\
kunz@theorie.physik.uni-oldenburg.de \\
${}^3$ ZARM, Universit\"at Bremen, Am Fallturm, 28359 Bremen, Germany \\
laemmerzahl@zarm.uni-bremen.de}

\maketitle

\begin{abstract}

The Schwarzschild--de Sitter space--time describes the gravitational field of a spherically symmetric mass in a universe with cosmological constant $\Lambda$. Based on this space--time we calculate Solar system effects like gravitational redshift, light deflection, gravitational time delay, Perihelion shift, geodetic or de Sitter precession, as well as the influence of $\Lambda$ on a Doppler measurement, used to determine the velocity of the Pioneer 10 and 11 spacecraft. For $\Lambda=\Lambda_0 \sim 10^{-52}\;{\rm m}^{-2}$ the cosmological constant plays no role for all of these effects, while a value of $\Lambda \sim - 10^{-37}\;{\rm m}^{-2}$, if hypothetically held responsible for the Pioneer anomaly, is not compatible with the Perihelion shift. 
\end{abstract}

\section{Introduction}

Until now all gravitational effects in the Solar system
and in binary systems are well described by means of the Einstein General Theory of Relativity \cite{Will93,Will01}\footnote{A possible exception is the Pioneer anomaly \cite{Andersonetal02}.}.
Cosmology, in contrast, is currently confronted with two mysteries:
these are dark matter and dark energy. 
Dark matter is introduced to obtain 
the gravitational field needed to describe observations
like the galactic rotation curves, gravitational lensing, 
or the structure of the cosmic microwave background.
Dark energy is envoked to explain the observed
accelerated expansion of the universe 
by means of an additional energy-momentum component.
Thus standard cosmology addresses the observed imbalance
in the gravitational field equations by postulating
unknown forms of matter and energy.

In general relativity, the repulsion necessary to obtain an accelerated expansion of the universe can be provided by the inclusion of a vacuum energy.
This corresponds to the well--known modification of the Einstein equations
consisting of the addition of a cosmological term $\Lambda g_{\mu\nu}$.
Observational data suggest for the cosmological constant $\Lambda$ a value of 
$\Lambda_0 \sim 10^{-52}\;{\rm m}^{-2}$ \cite{PeeblesRatra03}.
Alternatively, vacuum energy can be modeled by a dynamical field, 
such as quintessence \cite{Wetterich88,PeeblesRatra03}.

Instead of postulating unknown forms of matter and energy
another line of resolving these cosmological issues
consists of a modification of the gravitational field. 
The galactic rotation curves for a wide class of galaxies 
could be explained by using an additional gravitational Yukawa potential 
\cite{Sanders84}, 
for example\footnote{Another idea is to modify the Newtonian dynamics 
\cite{Milgrom02}.}. 
More recently brane--world models are considered as modified--gravity theories
in this context \cite{DvaliGabadadzePorrati00}.

We here consider the gravitational field of a spherically symmetric mass
in a universe with cosmological constant $\Lambda$,
described by the Schwarzschild--de Sitter space--time.
We calculate how the presence of such a cosmological constant
would affect Solar system effects
like gravitational redshift, light deflection, gravitational time delay,
geodetic or de Sitter precession. 
Comparing the theoretical results with the observations
gives Solar system estimates on the magnitude of such an
additional cosmological term in the Einstein equations.

Our investigation indicates that no present or future observation within the Solar system has the capability to reveal effects due to a cosmological constant 
with the current value of $\Lambda_0 \sim 10^{-52}\;{\rm m}^{-2}$ \cite{PeeblesRatra03}. We conclude that the cosmic acceleration therefore
does not lead to observable Solar system effects.

We then address the influence of a cosmological constant $\Lambda$, as present in the Schwarzschild--de Sitter space--time, on a Doppler measurement, used to determine the velocity of the Pioneer 10 and 11 spacecraft. By interpreting this $\Lambda$ not necessarily as the cosmological constant with value $\Lambda_0$, we note that a value of $\Lambda \sim - 10^{-37}\;{\rm m}^{-2}$ may lead to an anomalous acceleration on the order of the Pioneer acceleration. However, this value is not compatible with the Perihelion shift and, furthermore, leads to a position--dependent acceleration. 

We like to emphasize that the observational basis of the Pioneer anomaly is by far not of the same quality as the Solar system effects. It is still possible that the Pioneer anomaly is just a systematic effect, though no convincing explanation for this possibility has been found \cite{Andersonetal02}. Therefore, it is reasonable to take into account the possibility that this anomaly is due to some until now missed effect within standard physics, or that it even signals some kind of non--standard physics. Therefore we are looking for effects related to a space--time metric modified by a cosmological constant. This may also be of relevance for the design of a new space mission to test the long range gravitational field \cite{Dittusetal05}.

\section{Schwarzschild-de Sitter space--time}

We calculate Solar system effects for the spherically symmetric Schwarzschild--de Sitter metric \cite{Rindler01}
\begin{equation} \label{metric} 
ds^2 = \alpha(r) dt^2 - \alpha^{-1}(r) dr^2 - r^2 \left(d\theta^2 + \sin^2\theta d\phi^2\right)\, ,
\end{equation}
where 
\begin{equation}
\alpha(r) = 1 - \frac{2 M}{r} - \frac{1}{3} \Lambda r^2
\, , \end{equation}
and $\Lambda$ is the cosmological constant, and $M$ the mass of the source.
It is an exterior solution of the Einstein field equations
for a spherical mass distribution
\begin{equation}
R_{\mu\nu} - \frac{1}{2} g_{\mu\nu} R + \Lambda g_{\mu\nu} = 8 \pi G T_{\mu\nu} \, .
\end{equation}

In the following we take $\Lambda$ to be a free parameter
and try to obtain constraints on the magnitude of this constant 
from Solar system effects. 
Therefore our aim is twofold: 
First to investigate whether there might be some influence 
of the cosmological constant $\Lambda_0$ on Solar system effects,
and, second, to present estimates on the magnitude of such a constant $\Lambda$,
when it would not correspond to the cosmological constant, 
but instead would be of, e.g., galactic origin.   

The perihelion shift for the Schwarzschild--de Sitter space--time 
can be found in \cite{Rindler01} and \cite{KerrHauckMashhoon03}. 
Based on the Kerr--de Sitter metric in the latter paper 
\cite{KerrHauckMashhoon03} also gravitomagnetic effects have been calculated. 

\section{Gravitational redshift}

Since Schwarzschild--de Sitter is a stationary space--time there is a time--like Killing vector so that in coordinates adapted to the symmetry the ratio of the measured frequency of a light ray crossing different positions is given by
\begin{equation}
\frac{\nu}{\nu_0} = \sqrt{\frac{g_{00}(r)}{g_{00}(r_0)}} \, .
\end{equation}
For experiments in the vicinity of the central body (which in this case will be taken to be the Earth), we have $\Lambda r^2 \ll M/r$ so that to first order in $\Lambda$
\begin{equation}
\frac{\nu}{\nu_0} = \sqrt{\frac{g_{00}^{(0)}(r)}{g_{00}^{(0)}(r_0)}} \left(1 + \frac{\Lambda}{6} \left(\frac{r_0^2}{g_{00}^{(0)}(r_0)} - \frac{r^2}{g_{00}^{(0)}(r)}\right)\right) \, ,
\end{equation}
where $g_{00}^{(0)} = 1 - \frac{2 M}{r}$. For small masses $M$ this simplifies to
\begin{equation}
\frac{\nu}{\nu_0} = 1 - \frac{M}{r} + \frac{M}{r_0} - \frac{\Lambda}{6} \left(r^2 - r_0^2\right)   \, ,
\end{equation}
where we neglected products of $\Lambda$ and $M$. 

Clock comparison can be performed with an accuracy of $10^{-15}$, the H--maser in the GP--A redshift experiment \cite{VessotLevine80} reached a $10^{-14}$ accuracy. Since all observations are well described within Einstein theory, we conclude
\begin{equation}
|\Lambda| \lesssim 10^{-28}\;{\rm m}^{-2} \, ,
\end{equation} 
where we assumed a clock comparison between the Earth and a satellite at 15,000 km height. 

\section{Deflection of light}

The motion of point particles and light rays in the Schwarzschild--de Sitter space-time is described by the Lagrangian 
\begin{equation}\label{lagr} 
2 {\cal L} = \alpha(r) \dot{t}^2 - \alpha^{-1}(r) \dot{r}^2 - r^2 \dot{\phi}^2 \ ,
\end{equation}
where a dot stands for differentiation with respect to the affine parameter $\lambda$, and where from the very beginning the motion is restricted to the $\theta=\pi/2$ plane. 
Since the metric (\ref{metric}) is independent of $t$ and $\phi$ the quantities
\begin{equation}\label{const} 
E \equiv \alpha(r) \frac{dt}{d\lambda} \ , \qquad L \equiv r^2\frac{d\phi}{d\lambda}\ ,
\end{equation}
are conserved along the orbit of the particle and define its energy and angular momentum, respectively.

For light, the Lagrangian (\ref{lagr}) vanishes. Multiplication of (\ref{lagr}) by $\alpha/L^2$ and substitution of the angular momentum yields
\begin{equation}
\frac{d\phi}{dr} = \pm \frac{1}{r^2} \left[\frac{1}{b^2} - \frac{\alpha(r)}{r^2}\right]^{-1/2} \, , \label{dphidr}
\end{equation} 
where $b \equiv L/E$ which in Schwarzschild space--time can be interpreted as impact parameter. The $\pm$ is related to increasing/decreasing $r$. The distance of closest approach $r_0$ defined by
\begin{equation}
\left.\frac{dr}{d\phi}\right|_{r = r_0} = \frac{1}{b^2} - \frac{\alpha(r_0)}{r_0^2} = 0\, . \label{defr0}
\end{equation}
relates $b$ to $r_0$. When we replace $b$ by $r_0$ in (\ref{dphidr}) the constant $\Lambda$ drops out
\begin{equation}
\frac{d\phi}{dr} = \pm \frac{1}{r^2} \left[\frac{1}{r_0^2} \left(1 - \frac{2 M}{r_0}\right) - \frac{1}{r^2} \left(1 - \frac{2 M}{r}\right)\right]^{-1/2} \, .
\end{equation} 
Therefore the constant $\Lambda$ has no influence on the light deflection. This has already been recognized in \cite{Lake02}.

\section{Gravitational time delay}

We consider the standard measurement of the time delay of light where a radar signal is sent from the Earth to pass close to the Sun and reflect off another planet or a spacecraft. The time interval between the emission of the first pulse and the reception of the reflected pulse is measured. Let $r_\oplus$ and $r_R$ be the distance between the Sun and the Earth and the reflector, respectively. 

The time interval between emission and return of a pulse as measured by a clock on the Earth is 
\begin{equation} 
\Delta T = 2 t(r_\oplus, r_0) + 2 t(r_{R}, r_0) \, , 
\end{equation}
where $t(r_1, r_2)$ is the elapsed coordinate time along the signal's trajectory between distances $r_1$ and $r_2$ to the Sun. The calculation of $t(r_1, r_2)$ requires to represent $t$ as a function of $r$ along the path of the pulse. From (\ref{const}) and (\ref{dphidr}) we obtain
\begin{equation}
\frac{dt}{dr} = \pm \frac{1}{\alpha b} \left(\frac{1}{b^2} - \frac{\alpha(r)}{r^2}\right)^{-1/2}
\end{equation}
and, thus,
\begin{equation}
t(r, r_0) = \int^r_{r_0} \frac{dr}{b \alpha(r)} \left(\frac{1}{b^2} - \frac{\alpha(r)}{r^2}\right)^{-1/2}
\, , \label{timedelay}
\end{equation}
where we can replace $b$ by $r_0$ using (\ref{defr0}).

The result of this integration in the Schwarzschild space--time can be found, e.g., in~\cite{Hartle03}. The first order correction to (\ref{timedelay}) due to $\Lambda$ is
\begin{equation}
t_\Lambda(r, r_0) = \frac{\Lambda}{18} \left[\sqrt{r^2 - r_0^2}\left(2 r^2 + r_0^2\right)
+ 3 M \left(4 r \sqrt{r^2 - r_0^2} + r_0^2 \left(2 + \frac{\sqrt{r^2 - r_0^2}}{r + r_0}\right)\right)
\right] \, ,
\end{equation}
where terms of the order $M^2$ have been neglected. Since for Solar system observations  $r_0/r_R \ll 1$ and $r_0/r_\oplus \ll 1$ we obtain as $\Lambda$--induced modification of the time delay of a round trip of a signal
\begin{equation}
\Delta t_\Lambda = \frac{2 \Lambda}{9} \left(r_\oplus^3 + r_R^3 + 6 M(r_\oplus^2 + r_R^2) - 3 M r_0^2\right)\, .  
\end{equation}

In the recent Cassini experiment \cite{BertottiIessTortora03} one has not measured the time delay but, instead, the relative change in the frequency
\begin{equation}
y = \frac{\nu(t) - \nu_0}{\nu_0} = \frac{d}{dt} \Delta t \, , 
\end{equation}
where $\nu_0$ is the frequency of the radio waves emitted on Earth. These radio waves reach Cassini with a frequency $\nu^\prime$ and are sent back with the same frequency $\nu^\prime$. The frequency of this signal reaching the Earth is $\nu(t)$. The contribution to the frequency change from the $\Lambda$--term is
\begin{equation}
y_\Lambda = \frac{d}{dt} \Delta t_\Lambda = - \frac{4}{3} \Lambda M r_0(t) \frac{d r_0(t)}{dt} \, ,
\end{equation}
where for the Cassini setup $dr_0/dt \approx v_\oplus$ is approximately the velocity of the Earth orbiting the Sun. That means that the $\Lambda$--contribution to the signal increases with distance. The Schwarzschild contribution to $y$ is $6 \cdot 10^{-10}$ and has been verified by the Cassini measurements within an accuracy of $y$ of $\sim 10^{-14}$. 

Cassini measured the time delay for approximately 25 days, that is, during the period of 12 days before and 12 days after conjunction \cite{BertottiIessTortora03}. During one day the distance of closest approach of the signal changed by $\sim 1.5$ Solar radii. Therefore, the measured signal would be
\begin{equation}
y_\Lambda(12 {\rm d}) - y_\Lambda(0) = 16 \Lambda M_\odot R_\odot v_\oplus \, .
\end{equation}
Since the Cassini measurement was in accordance to the Einstein prediction, we conclude from $y_\Lambda(12 {\rm d}) - y_\Lambda(0) \leq 10^{-14}$ that 
\begin{equation}
|\Lambda| \leq \frac{10^{-14} c^3}{16 G M_\odot R_\odot v_\oplus} \approx 10^{-24} \; {\rm m}^{-2}\, .
\end{equation}

\section{Perihelion shift}

The perihelion shift is based on the modification of the geodesic equation for a massive particle which can be derived in a standard manner. In first order in $\Lambda$ and with $u = 1/r$ one obtains 
\begin{equation}
\frac{d^2 u}{d\varphi^2} = \frac{M}{L^2} - u + 3 M u^2 - \frac{\Lambda}{3 L^2 u^2}
\end{equation}
which leads to an additional shift \cite{KerrHauckMashhoon03}
\begin{equation}
\delta = \pi \Lambda a^2 \frac{a}{M} \sqrt{1 - e^2} = \Delta \frac{\Lambda a^4 (1 - e^2)^{3/2}}{6 M^2} 
\end{equation}
where $e$ is the eccentricity of the orbit 
and $\Delta$ the standard Einstein expression for the perihelion shift, see also \cite{Rindler01,Islam83,KraniotisWhitehouse03}. 
(Obviously, for a smaller central mass $M$, 
implying a smaller Perihelion shift, 
the influence of the $\Lambda$--term becomes more important. 
Its effect should be more pronounced for the outer planets 
and for comets with large $a$.)
The perihelion shift of Mercury, $43^{\prime\prime}$ per century, 
is in full agreement with Einstein theory 
within the accuracy of 430 $\mu$as \cite{Nordtvedt01a}. 
From that we conclude\footnote{After submission of this paper similar analyses appeared \cite{Iorio05b,JetzerSereno06}.}
\begin{equation}
|\Lambda| \leq \frac{430\; \mu{\rm as}}{\Delta} \frac{6 M^2}{a^4 (1 - 4 e^2)} \approx 10^{-41}\;{\rm m}^{-2} \, . 
\end{equation}

\section{Geodetic precession}\label{SecGeodPrec}

We calculate the geodetic precession by introducing a coordinate system rotating with angular velocity $\omega$. The new angular coordinate $\varphi$ is defined through
\begin{equation}
d\varphi = d\phi - \omega dt  \, , 
\end{equation}
and gives in the equatorial plane $\theta=\pi/2$ the metric
\begin{equation} \label{metric_changed}
ds^2 = \left(\alpha - r^2 \omega^2\right) \left(dt - \frac{r^2\omega}{\alpha - r^2\omega^2} d\varphi\right)^2 - \alpha^{-1} dr^2 - \frac{r^2\alpha}{\alpha-r^2\omega^2} d\varphi^2 \, . 
\end{equation}
The canonical form of the metric is
\begin{equation}
\label{canon_metric}
ds^2 = e^{2\Psi} \left(dt - w_i dx^i\right)^2 - h_{ij} dx^i dx^j \ ,
\end{equation}
with
\begin{align}
e^{2\Psi} & = \alpha - r^2 \omega^2 \ , & \qquad w_1 & = w_2 = 0 \, , & \qquad w_3 & = \frac{r^2\omega}{\alpha - r^2\omega^2} \\
h_{11} & = \alpha^{-1} \ , & \qquad h_{33} & = \frac{r^2\alpha}{\alpha-r^2\omega^2} \ ,
 & &  \nonumber 
\end{align}
with the indices $1, 2, 3$ referring to $r, \theta, \varphi$,
respectively. For a given $r = r_0$ we can choose an $\omega$ so that $\partial \Psi/\partial r = 0$. Since $\Psi$ is related to the acceleration, this means that the line $r = r_0$ is geodesic. This
corresponds to a modified Kepler frequency
\begin{equation}\label{omega_prec} 
\omega^2 = \frac{M}{r^3} - \frac{\Lambda}{3} \, , 
\end{equation}
also obtained in \cite{KerrHauckMashhoon03}, and implies
\begin{align}
e^{2\Psi} & = 1 - \frac{3M}{r} \ , & \qquad w_1 & = w_2 = 0 \, , & \qquad w_3 & = \frac{r^2\omega}{1 - 3 M/r} \\
h_{11} & = \alpha^{-1} \ , & \qquad h_{33} & = \frac{r^2\alpha}{1 - 3 M/r} \, .
 & &  \nonumber 
\end{align}

The rotation rate of a gyroscope at a fixed position in the rotating coordinate system is given by \cite{Rindler01}
\begin{equation}
\Omega^2 = \frac{e^{2 \Psi}}{8} h^{ik} h^{jl} \left(w_{i,j} - w_{j,i}\right) \left(w_{k,l} - w_{l,k}\right) 
\end{equation}
and points in the negative direction relative to the orbit. We obtain
\begin{equation}
\Omega = \sqrt{\frac{M}{r^3} - \frac{\Lambda}{3}} = \Omega_0 \left(1 - \frac{\Lambda r^3}{6 M}\right) \, , \qquad \Omega_0 = \sqrt{\frac{M}{r^3}} \, . 
\end{equation}
For an alternative derivation of this geodetic precession, see the Appendix.

The mission GP-B will measure the geodetic precession $\Omega_0 = 6,600\; {\rm mas/y}$ with an expected accuracy of 0.1 mas/y. If the value predicted from Einstein theory will be confirmed and no additional term is seen, we are led to an estimate for $\Lambda$
\begin{equation}
|\Lambda| \leq 10^{-27} \; {\rm m}^{-2} \ .
\end{equation}

\section{Doppler tracking of satellites on escape trajectories}

We now calculate the two--way Doppler tracking of a satellite moving on a geodesic radially away from the Sun. 
This amounts to an investigation of whether the presence of a cosmological constant might be made responsible 
for the anomalous acceleration observed for the Pioneer spacecraft \cite{Andersonetal02}. For simplicity, we take an observer at a fixed radial distance from the Sun and a spacecraft moving radially so that Sun, Earth and spacecraft lie on one line. 

The measurement is carried out as follows: 
On Earth a radio signal with frequency $\nu_0$ 
is sent to the spacecraft at position $r$,
where it will arrive redshifted according to 
\begin{equation}
\tilde\nu = \sqrt{\frac{\alpha(r)}{\alpha(r_\oplus)}} \nu_0 \, .
\end{equation}
Due to the velocity of the spacecraft the measured frequency 
is in addition Doppler shifted
\begin{equation}
\nu^\prime = \frac{1}{\sqrt{1 - (v^r)^2}} \left(1 - v^r\right) \tilde\nu \, ,
\end{equation}
where $v^r = \theta^r(v)$ is the measured radial velocity 
($v$ is the 4--velocity of the spacecraft 
and $\theta^r$ is the normalized basis 1--form in $r$--direction, 
$g^{\mu\nu} \theta^r_\mu \theta^r_\nu = -1$; 
in coordinates, $\theta^r_\mu = \alpha^{-1/2} (0, 1, 0, 0)$). 
This is also the frequency sent back to Earth, where it arrives with the frequency
\begin{eqnarray}
\nu^{\prime\prime} = \frac{1}{\sqrt{1 - (v^r)^2}} \left(1 - v^r\right) \sqrt{\frac{\alpha(r_\oplus)}{\alpha(r)}} \nu^\prime \, . 
\end{eqnarray}
The two--way Doppler tracking will then yield
\begin{equation}
\frac{\nu^{\prime\prime} - \nu_0}{\nu_0} = - 2 \frac{v^r}{1 + v^r} \approx - 2 v^r \, . 
\end{equation}
As expected, the gravitational redshift drops out. 
Only the motion of the spacecraft influences the result. 

Therefore, what is finally needed is the solution for the radial velocity $v$ 
obtained from the geodesic equation
\begin{eqnarray}
0 & = & \frac{d^2 r}{dt^2} + \left(\Bigl\{\overset{r}{{}_{\mu\nu}}\Bigr\} - \Bigl\{\overset{0}{{}_{\mu\nu}}\Bigr\} \frac{dr}{dt}\right) \frac{dx^\mu}{dt} \frac{dx^\nu}{dt} \\
& = & \frac{d^2 r}{dt^2} - \frac{3}{2} \frac{1}{\alpha} \partial_r \alpha \left(\frac{dr}{dt}\right)^2 + \frac{1}{2} \alpha \partial_r \alpha \, ,
\end{eqnarray}
where $\left\{\overset{\mu}{{}_{\rho\sigma}}\right\}$ are the Christoffel symbols. 
We can safely neglect the term quadratic in the velocities. 
Then, to first order in $M$ and $\Lambda$ we obtain
\begin{equation}
\frac{d^2 r}{dt^2} = - \frac{M}{r^2} + \frac{\Lambda}{3} r \, . 
\end{equation}
This leads to
\begin{equation}
\frac{1}{2} \left(\frac{d r}{dt}\right)^2 = E + \frac{M}{r} + \frac{1}{6} \Lambda r^2 \, ,
\end{equation}
where $E$ is an integration constant. 

The Doppler measurement in terms of the measured velocity is then
\begin{equation}
\frac{\nu^{\prime\prime} - \nu_0}{\nu_0} = - 2 \frac{v^r}{1 + v^r} = - 2 \frac{dr/dt}{\sqrt{\alpha(r)} + dr/dt}  
\end{equation}
We assume $\Lambda r^2, M/r \ll E$ 
and make an expansion with respect to $\Lambda$ and with respect to $M$. 
Furthermore, 
\begin{equation}
\frac{\nu^{\prime\prime} - \nu_0}{\nu_0} = -2 \frac{\sqrt{2 E}}{1 + \sqrt{2 E}} - \frac{2 M}{r} \frac{1 + 2 E}{(1 + \sqrt{2 E})^2 \sqrt{2 E}} - \frac{1}{3} \Lambda r^2 \frac{1 + 2 E}{(1 + \sqrt{2 E})^2 \sqrt{2 E}} + \mathcal{O}(\Lambda^2, M^2)
\end{equation}
Since $2 E$ corresponds to $v_0^2/c^2$, the non--relativistic limit is
\begin{equation}
\frac{\nu^{\prime\prime} - \nu_0}{\nu_0} = -2 \sqrt{2 E} - \frac{2 M}{r} \frac{1}{\sqrt{2 E}} - \frac{1}{3} \Lambda r^2 \frac{1}{\sqrt{2 E}} + \mathcal{O}(\Lambda^2, M^2)
\end{equation}
The first two terms are the standard Newtonian contributions. 
The position $r$ in the $\Lambda$--term 
induces an additional time--dependence of the two--way Doppler signal. 
Approximating $r = r_0 + v_0 t$ and $E = v_0^2/2$, 
the time--dependent signal related to the $\Lambda$--term is
\begin{equation}
\frac{d}{dt} \left(\frac{\nu^{\prime\prime} - \nu_0}{\nu_0}\right)_\Lambda = - \frac{2}{3} \Lambda r \dot r \frac{1}{\sqrt{2 E}} = - \frac{2}{3} \Lambda (r_0 + v_0 t) \label{PioneerDrift}
\end{equation}
If the first term is assumed to be responsible for the observed Pioneer anomaly, 
then with $r_0$ being on the order of $100\;{\rm AU}$ 
the constant $\Lambda$ should be on the order of $- 10^{-37}\;{\rm m}^{-2}$. 
This value is not compatible with the Perihelion shift. That means that there is no way to simulate the Pioneer anomaly with a term of the form of a cosmological constant without being in conflict with observations. 

The influence of the cosmological constant on the semi--major axis is negligible. For any non--Newtonian gravitational force the distance $r$ of the planetary orbits as function of the angle $\varphi$ is given by 
\begin{equation}
r(\varphi) = \frac{r_0 (1 + \epsilon)}{1 + e \cos((1 + \epsilon) \varphi)} 
\end{equation}
where $\epsilon$ is the parameter related to the non--Newtonian part of the gravitational interaction describing the perihelion shift and the change in the planetary orbital parameter $r_0$ (which is related to the semi--major axis). As a consequence, we can relate the perihelion shift to the change in the orbital parameter: If, due to some non--Newtonian or non--Einsteinian gravitational forces, there is an additional perihelion shift of the order $400 \mu{\rm as}$ per century (which is the present day accuracy \cite{Nordtvedt01a}), then $\epsilon \approx 10^{-12}$ so that $\epsilon r_0 \approx 15\;{\rm cm}$. This is clearly beyond any observational access for any planet in the Solar system. Therefore, a value of $\Lambda$ which may give for the Pioneers an acceleration of the required order has practically no influence on the planetary orbits. 


\section{Conclusion and outlook}

We have calculated the Solar system effects 
in the frame of the Schwarzschild--de Sitter space--time,
treating the cosmological constant $\Lambda$
in the metric as a free parameter.
Comparison with observations then leads to the following 
constraints on the magnitude of the constant $\Lambda$:

\begin{table}[h]
\begin{center}
\begin{tabular}{ll}
Observed effect & Estimate on $\Lambda$ \\ \hline
gravitational redshift & $|\Lambda| \leq 10^{-27}\;{\rm m}^{-2}$ \\ 
perihelion shift & $|\Lambda| \leq 10^{-41}\;{\rm m}^{-2}$ \\
light deflection & no effect  \\
gravitational time delay & $|\Lambda| \leq 6 \cdot 10^{-24}\;{\rm m}^{-2}$ \\
geodetic precession & $|\Lambda| \leq 10^{-27}\;{\rm m}^{-2}$ \\
Pioneer anomaly & $\Lambda \sim - 10^{-37}\;{\rm m}^{-2}$ 
\end{tabular}
\end{center}
\caption{Estimates on $\Lambda$ from Solar system observations.}
\end{table} 

It is remarkable that the perihelion shift is by far the most sensitive tool to observe an influence of the cosmological constant on the scale of he Solar system which is probably due to the fact that it constitutes a cumulative long term effect. For light deflection the presence of $\Lambda$ has no effect.

Assuming that $\Lambda$ represents the cosmological constant
with the current value of $\Lambda_0 \sim 10^{-52}\;{\rm m}^{-2}$,
we conclude, that no present or future observation of these Solar system effects
has the capability to reveal the presence of such a cosmological constant.
This also extends to Solar system effects 
like the gravitomagnetic clock effect or time delay,
based on the influence of an additional gravitomagnetic field 
\cite{KerrHauckMashhoon03}. 

Addressing the influence of a cosmological constant $\Lambda$
on a Doppler measurement, used to determine
the velocity of the Pioneer 10 and 11 spacecraft,
we conclude that a value of $\Lambda \sim - 10^{-37}\;{\rm m}^{-2}$
may lead to an acceleration of the order of the 
observed anomalous Pioneer acceleration.
However, the acceleration caused by $\Lambda$ changes with distance and is in conflict with the observed Perihelion shift. Therefore the hypothetical presence of a 
negative cosmological constant $\Lambda \sim - 10^{-37}\;{\rm m}^{-2}$
as a possible explanation for the Pioneer anomaly has to be excluded.

We consider the calculation of Solar system effects 
in the Schwarzschild--de Sitter space--time 
only as a first step towards the general goal,
of obtaining observational constraints from Solar system effects 
for modified theories of gravity.
We envisage to repeat this kind of calculation in more general contexts
like quintessence \cite{Wetterich88,PeeblesRatra03} 
(for a first step in this direction see \cite{Mbelek04}), 
in varying $G$ scenarios \cite{ReuterWeyer04}, 
in dilaton scenarios \cite{DamourPiazzaVeneziano02,DamourPiazzaVeneziano02a}, 
and in braneworld models \cite{DvaliGabadadzePorrati00,Lue05} (see \cite{Iorio05,Iorio05a} for the discussion of one related effect.). 

\section*{Acknowledgement}

We would like to thank H. Dittus and B. Mashhoon for valuable discussions. 
V.K. thanks the Deutsche Forschungsmeneinschaft DFG for financial support. 
C.L. acknowledges financial support of the German Space Agency DLR.

\section*{Appendix: Alternative derivation of geodetic precession}

A gyroscope is a spin four-vector $s^\mu$ which moves along a timelike geodesic with four-velocity $u^\mu$. The spin is space--like 
\begin{equation}\label{veloc_spin} 
g_{\mu\nu} u^\mu s^\nu = 0 \ .
\end{equation}
and is paralelly propagated along $u^\mu$
\begin{equation}\label{geodesic}
\frac{ds^\mu}{d\lambda} + \Bigl\{\overset{\mu}{{}_{\nu\rho}}\Bigr\} s^\nu u^\rho = 0 \, .
\end{equation}
An observer moving with the gyroscope on a circular orbit in the equatorial plane will see its spin precess in the equatorial plane, see e.g. \cite{Hartle03}. 

The components of the four--velocity of the gyroscope in the $\theta=\pi/2$ plane are given by $u^\mu = u^{t}(1,0,0,\omega)$ with $\omega$ from (\ref{omega_prec}) and $u^t$ is determined by $g_{\mu\nu} u^\mu u^\mu = 1$. Suppose that the spin initially points in the equatorial plane along the unit vector $e_{\hat r}^\mu = (0, \alpha^{1/2}, 0, 0)$ in $r$--direction. From (\ref{geodesic}) $ds^\theta/d\lambda = 0$ so that an initial $s^\theta = 0$ remains zero. 
From (\ref{veloc_spin}) we obtain
\begin{equation}
s^{t}=r^2\omega\alpha^{-1}s^{\phi} \, , 
\end{equation}
and (\ref{geodesic}) gives
\begin{eqnarray}
\label{s_r}
\frac{ds^{r}}{dt}+(r-3M)\omega s^{\phi} = 0 \ , \\
\label{s_phi}
\frac{ds^{\phi}}{dt}-\frac{\omega}{r}s^{r} = 0 \, . 
\end{eqnarray}
Choosing the initial condition $s^\phi(0)=0$ and introducing
\begin{equation}
\Omega = \left(1 - \frac{3M}{r}\right)^{1/2} \omega = \sqrt{\left(1 - \frac{3M}{r}\right) \left(\frac{M}{r^3} - \frac{\Lambda}{3}\right)} \approx \sqrt{\frac{M}{r^3} - \frac{\Lambda}{3}}
\end{equation}
the solution of (\ref{s_r}) and (\ref{s_phi}) is
\begin{eqnarray}\label{s_r1} 
s^r & = & s_\ast \alpha^{1/2}\cos{(\Omega t)} \\
\label{s_phi1}
s^\phi & = & s_\ast \alpha^{1/2} \left(\frac{\omega}{\Omega r}\right)\sin{(\Omega t)}
\, .
\end{eqnarray}
where $s_\ast =(- s^\mu s_\mu)^{1/2}$. 



\end{document}